\documentclass[11pt,longnamesfirst,preprint]{aastex}
\received{2003 November~18}
\revised{2004 April~7}
\accepted{2004 May~4}
\paperid{59362}

\shorttitle{Extrasolar Planet Radio Emission}
\shortauthors{Lazio et al.}

\begin{document}
\title{The Radiometric Bode's Law and Extrasolar Planets}

\author{T.~Joseph~W.~Lazio}
\affil{Naval Research Laboratory, Code~7213, Washington, DC
	20375-5351}
\email{Joseph.Lazio@nrl.navy.mil}

\author{W.~M.~Farrell}
\affil{NASA Goddard Space Flight Center, Code~695, Greenbelt, MD 20771}
\email{william.m.farrell@gsfc.nasa.gov}

\author{Jill Dietrick, Elizabeth
	Greenlees, Emily Hogan, Christopher Jones, L.~A.~Hennig}
\affil{Thomas Jefferson High School for Science and Technology, 6650 
	Braddock Road, Alexandria, VA  22312}
\email{hennig@lan.tjhsst.edu}

\begin{abstract}
We predict the radio flux densities of the extrasolar planets in the
current census, making use of an empirical relation---the radiometric
Bode's Law---determined from the five ``magnetic'' planets in the
solar system (\objectname[]{Earth} and the four gas giants).  Radio
emission from these planets results from solar-wind powered electron
currents depositing energy in the magnetic polar regions.  We find
that most of the known extrasolar planets should emit in the frequency
range 10--1000~MHz and, under favorable circumstances, have typical
flux densities as large as 1~mJy.  We also describe an initial,
systematic effort to search for radio emission in low radio frequency
images acquired with the Very Large Array.  The limits set by the VLA
images ($\approx 300$~mJy) are consistent with, but do not provide
strong constraints on, the predictions of the model.  Future radio
telescopes, such as the Low Frequency Array (LOFAR) and the Square
Kilometer Array (SKA), should be able to detect the known extrasolar
planets or place austere limits on their radio emission.  Planets with
masses much lower than those in the current census will probably
radiate below~10~MHz and will require a space-based array.
\end{abstract}

\section{Introduction}\label{sec:intro}

The past few years have been an exciting time as
extrasolar planets have been demonstrated to be widespread and multiple
planetary systems have been found.  The current census now numbers
more than 100 extrasolar planets, in over 90 planetary systems
\citep{s03,m03}.

The vast majority of these extrasolar planets have been detected via
the reflex motion of the host star.  As the existing census shows,
this method has proven to be wildly successful.  Nonetheless, the
reflex motion of the star is a measure of the planet's gravitational
influence and is necessarily an \emph{indirect} detection of the
planet.  As a consequence, the only property of the planet that one
can infer is its mass, and because of the mass function's dependence
on the inclination angle ($\sin i$), one can infer only a minimum
mass.

Direct detection of reflected, absorbed, or emitted radiation from a
planet allows for the possibility of complementary information, and
likely a more complete characterization of the planet.  The prototype
of such a direct detection is the detection of sodium absorption lines
in the atmosphere of the planet orbiting HD~209458 \citep{cbng02}.
Unfortunately, the incidence of transiting planets will always remain
low relative to the total number of planets known.

The \objectname[]{Earth} and gas giants of our solar system are
described commonly as ``magnetic planets'' because they contain
internal dynamo currents that are capable of generating a
planetary-scale magnetic field.  The dynamo currents themselves arise
from the rapid rotation of a conducting fluid.  The composition of the
fluid varies from planet to planet, being liquid iron in the
\objectname[]{Earth}, probably metallic hydrogen in
\objectname[]{Jupiter} and \objectname[]{Saturn}, and probably a salty
ocean in \objectname[]{Uranus} and \objectname[]{Neptune}.  In turn,
these planetary magnetic fields are immersed in the high-speed
electrical gas emitted by the \objectname[]{Sun}.  Via a coupling
between the solar wind and the planetary magnetic field, all of these
magnetic planets produce radio emission.

Some of the extrasolar planets are also magnetic planets.
\cite{swb03} have detected a modulation in the \ion{Ca}{2}~H and~K
lines of \objectname[HD]{HD~179949} with a periodicity which is that
of the planetary orbit.  They interpret this as a magnetic interaction
between the star and planet, though there is no constraint as yet on
the magnetic field strength of the planet.  Other stars in their
sample also appear to have modulations in the \ion{Ca}{2}~H and~K
lines, though it is not yet clear that these modulations are periodic
with the orbital period of their planets, as is the case for
\objectname[HD]{HD~179949}.

By analogy to solar system planets, there have been a number of
suggestions and searches for natural radio emission from extrasolar
planets.  Early searches were conducted by \cite{yse77} and
\cite{l91}.  Due to the lack of any known extrasolar planets at the
time, these were necessarily blind searches.  More recently,
\cite{fdz99} and \cite{ztrr01}, building on empirical relations
derived for solar system planets and 
calibrated by spacecraft fly-bys, have made specific predictions for
the radio emission from the known extrasolar planets.

Detection of radio emission from an extrasolar planet would constitute
a \emph{direct} detection and can yield fundamental information about
the planet.  First, a measurement of the radio emission is directly
indicative of the polar magnetic field strength at the planet.  For
example, the high-frequency cutoff of Jovian decametric bursts
($\simeq 40$~MHz) is interpreted as being due to the Jovian polar
magnetic field strength, which allowed an estimate for the strength of
the Jovian magnetic field nearly~20~yr prior to the first \textit{in
situ} magnetic field observations.  In turn, of course, the presence
of a magnetic field provides a rough measure of the composition of the
planet, insofar as it requires the planet's interior to have a
conducting fluid.  Combined with an estimate of the planet's mass, one
could deduce the composition of the fluid by analogy to the solar
system planets (liquid iron vs.\ metallic hydrogen vs.\ salty ocean).

Second, the periodic nature of the radio emission has been used to
define precisely the planetary rotation periods of all of the gas
giant planets in the solar system.  As the magnetic field is presumed
to be tied to the interior of the planet, it provides a more accurate
measure of the planet's rotation rate than atmospheric phenomena such
as clouds.  For instance, the rotation period of \objectname[]{Neptune} was
determined initially by observations of the differentially rotating
cloud tops but then was redefined after detection of the Neptunian
radio emission \citep{lzde93}.

Finally, testing the extent to which solar-system models of magnetic
fields can be applied to extrasolar planets may have important implications
for assessing the long-term ``habitability'' of any terrestrial
planets found in the future.  The importance of a magnetic field is
that it deflects incident cosmic rays.  If these particles reach the
surface of an otherwise habitable planet, they may cause severe
cellular damage and disruption of genetic material to any life on its
surface or may prevent life from arising at all.  A secondary importance
of a magnetic field is that it can prevent the planet's atmosphere
from being eroded by the stellar wind \citep{mitchelletal01}; this
process is thought to be a contributing factor to the relative
thinness of the Martian atmosphere.  (This is of course unlikely to be
an issue for extrasolar giant planets.)

This paper extends and updates initial attempts to predict the radio
emission of extrasolar giant planets by applying the solar-system
determined ``radiometric Bode's Law'' to the current census of
extrasolar planets.  In \S\ref{sec:law} we summarize briefly the
various forms of the radiometric Bode's Law and the form we have
adopted for use, in \S\ref{sec:census} we summarize the current state
of the extrasolar planet census and predict the ``burst'' emission levels
and frequencies for radio emission from these planets, in
\S\ref{sec:observe} we use the initial observations from a 74~MHz sky
survey to confront the model predictions with data, and in
\S\ref{sec:conclude} we assess how existing and future radio
telescopes may be able to detect radio emission from extrasolar
planets and present our conclusions.

\section{The Radiometric Bode's Law}\label{sec:law}

All of the planets in the solar system with substantial magnetic
fields are strong emitters of coherent cyclotron emission at radio
wavelengths.  The emission itself arises from energetic (keV)
electrons propagating along magnetic field lines into active auroral
regions \citep{g74,wl79}.  The source of the electron acceleration to
high energies is ultimately a coupling between the incident solar wind
and the planet's magnetic field.  The precise details are not well
understood, but the coupling arises from magnetic field reconnection
in which the magnetic field embedded in the solar wind and the
planetary magnetic field cancel at their interface, thereby energizing
the plasma.  Energetic electrons in the energized plasma form a
current flow planet-ward along the planet's magnetic field lines, with
the lines acting effectively like low resistance wires.  The energy in
these magnetic field-aligned electric currents is deposited in the
upper polar atmosphere and is responsible for the visible aurora.
Besides auroral emissions at visible wavelengths (e.g., northern
lights), about~$10^{-5}$--$10^{-6}$ of the solar wind input power is
converted to escaping cyclotron radio emission \citep{g74}.  The
auroral radio power from \objectname[]{Earth} is of order $10^8$~W
while for \objectname[]{Jupiter} it is of order $10^{10.5}$~\hbox{W}.

The cyclotron emission process described is generic, and typically
most efficient in the auroral region, about~1--3 planetary radii in
altitude.  Specific details of the cyclotron emission process do
differ from planet to planet, depending upon secondary effects such as
the polar source location and beaming angles, which in turn depend
upon the planet's magnetic field topology and low-altitude atmospheric
density.  Nonetheless, applicable to all of the magnetic planets is a
macroscopic relationship called the ``radiometric Bode's law''
relating the incident solar wind power, the planet's magnetic field
strength, and the median emitted radio power.  Various investigators
\citep{db84,dk84,dr85,bgd86,r87,d88,mg88} have found
\begin{equation}
P_{\mathrm{rad}} = \epsilon P_{\mathrm{sw}}^x,
\label{eqn:law}
\end{equation}
where $P_{\mathrm{rad}}$ is the median emitted radio power,
$P_{\mathrm{sw}}$ is the incident solar wind power, $\epsilon \sim
10^{-5}$--$10^{-6}$ is the efficiency at which the solar wind power is
converted to emitted radio power, and $x \approx 1$.  The values of
the coefficient~$\epsilon$ and the power-law index~$x$ measured by
various investigators have differed over the years based on the
quantity of data available; the early determinations were based on
fly-bys of only \objectname[]{Jupiter} and \objectname[]{Saturn} and
measurements of the \objectname[]{Earth's} radio emission while later
determinations include data from the Voyager~2 \objectname[]{Uranus}
and \objectname[]{Neptune} fly-bys as well.  In general, the more
recent work indicates higher efficiencies than the earlier work \citep{zarkaetal97}.

\cite{fdz99} and \cite{ztrr01} extrapolated this relation to
extrasolar planetary systems by making use of two scaling relations
and adopting more recent estimates for~$\epsilon$ and~$x$.  The
incident solar wind power depends upon the ram pressure of the solar
wind and the cross-sectional area presented by the magnetosphere to
the solar wind.  The ram pressure of the solar wind depends on the
star-planet distance~$d$ while the size of the planet's magnetosphere
is related to its magnetic moment.  \cite{b47} demonstrated that
planetary-scale, dynamo-driven magnetic fields are scalable, with a
functional dependence on body mass, radius, and angular frequency; and
the presence of planetary-scale magnetic fields around the gas giants
shows that his analysis does not depend upon the actual composition of
the conducting fluid in the planet's interior.  Blackett's law relates
a planet's magnetic moment (its surface field times its radius cubed,
$BR^3$) to its rotation rate and mass by
\begin{equation}
\mu \sim \omega M^2,
\label{eqn:blackett}
\end{equation}
where $\omega$ is the planet's rotation rate and~$M$ is its mass.
Since that time and with the inclusion of the Voyager magnetic field
observations at the outer planets, numerous planetary magnetic field
scaling laws have been derived, each expanding upon but maintaining
some similarity to Blackett's Law.

Combining these various scaling relations, \cite{fdz99} and
\cite{ztrr01} found that the planet's median radio power driven by the
stellar wind is
\begin{equation}
P_{\mathrm{rad}}
 \sim 4 \times 10^{11}\,\mathrm{W}\left(\frac{\omega}{10\,\mathrm{hr}}\right)^{0.79}\left(\frac{M}{M_J}\right)^{1.33}\left(\frac{d}{5\,\mathrm{AU}}\right)^{-1.6}.
\label{eqn:rbl}
\end{equation}
We have normalized all quantities to those of \objectname[]{Jupiter}.
\citet[Appendix~A]{fdz99} discussed slight differences to this
radiometric Bode's law, the differences result from the statistical
spread in the various (solar system) planets' magnetic moments and
amount to slightly different exponents and/or a different coefficient.

We stress various aspects of this radiometric Bode's law.  First, it
is grounded in \textit{in situ} measurements from spacecraft fly-bys
of the gas giants as well as measurements of the
\objectname[]{Earth's} radio emission.  In particular, \cite{fdz99}
considered two forms of the radiometric Bode's Law, one incorporating
measurements of only \objectname[]{Jupiter} and \objectname[]{Saturn}
and the other incorporating all five magnetic planets.  The former
gives lower power levels (by a factor of~100); in
equation~(\ref{eqn:rbl}), we have adopted the latter because it is
based on a complete census of magnetic planets in the solar system.

Second is the importance of the planet-star distance.  As
\cite{ztrr01} show, the radio power from \objectname[]{Earth} is
larger than that from \objectname[]{Uranus} or \objectname[]{Neptune}
even though both of those planets have magnetic moments approximately
50~times larger than that of the \objectname[]{Earth}.

Third, the power levels estimated by equation~(\ref{eqn:rbl}) are for
emission into a solid angle of~4$\pi$.  Emission solid angles for
solar system planets span a large range.  Considering Jovian
decametric emission at macroscopic scales, emission occurrence is as
$\Omega \sim \pi$.  In estimating the flux density, the relevant
quantity is $P/\Omega$.  (See equation~\ref{eqn:fluxdensity} below and
the discussion in \citealt{fdz99}.)  Assuming a smaller emission solid
angle would mean reducing the nominal power levels of
equation~(\ref{eqn:rbl}) by the factor $\Omega/4\pi$.  (This factor
also explains the difference between the nominal power level for
Jupiter predicted by equation~\ref{eqn:rbl} and the level quoted in
\S\ref{sec:intro}.)

Fourth, the radiometric Bode's Law of equation~(\ref{eqn:rbl})
describes the \emph{median} emitted power from the magnetospheres of
the \objectname[]{Earth} and all of the solar gas giants, including
the \emph{non-Io driven} Jovian decametric radio emission
\citep{ztrr01}.  Planetary magnetospheres tend to act as
``amplifiers'' of the incident solar wind, so that an increase in the
solar wind velocity (and therefore incident pressure) leads to
geometrically higher emission levels.  This effect is observed at all
of the magnetized planets.  For instance, the Saturnian radio flux has
been observed to increase by factors of~20 due to changes in the solar
wind velocity \citep{dr85}.  At the \objectname[]{Earth}, the radio
power has been observed to increase by a factor of~10 for every
200~km~s${}^{-1}$ increase in the solar wind velocity
\cite[Figure~5]{gd81}; the observed range of solar wind velocities at
the \objectname[]{Earth} is at least 400~km~s${}^{-1}$
(from~400~km~s${}^{-1}$ to~800~km~s${}^{-1}$).  The radio power levels
from a planet can exceed that predicted by equation~(\ref{eqn:rbl}) by
factors of~100 and possibly more.

Finally, an implicit assumption in our analysis is that the mass loss
rates for stars hosting planets are similar to that of the Sun.  The
stellar wind pressure incident on a planetary magnetosphere is
$P_{\mathrm{sw}} \propto \rho V^3$, where $\rho$ is the density of the
stellar wind and~$V$ is its velocity.  Stars with higher (lower) mass
loss rates than the \objectname[]{Sun} will have higher (lower)
stellar wind densities and should drive higher (lower) planetary radio
emissions.  Differences in the mass loss rate from star to star could
lead to statistical variations in the power levels we predict, though
these should be relatively less important than variability in stellar
wind velocities.  (Work is in progress to utilize what is known about
the various stellar ages and activities to relax this assumption.)

In order for gyrating electrons to generate radio emission that
escapes the auroral region, the Doppler-shifted electron cyclotron
frequency must couple directly to a magnetoplasma mode. This coupling
can only occur in low density, high magnetic field conditions where
the local electron plasma-to-cyclotron frequency ratio is much less
than one. The regions just above the polar ionosphere satisfy this
coupling criteria.  Hence, the characteristic emission frequency can
be related to the planetary magnetic dipole moment
(equation~\ref{eqn:blackett}) by
\begin{equation}
\nu_c
 \sim 23.5\,\mathrm{MHz}\,\left(\frac{\omega}{\omega_J}\right)\left(\frac{M}{M_J}\right)^{5/3}R_J^3.
\label{eqn:fc}
\end{equation}
In order to be detected from ground-based telescopes, $\nu_c$ must
exceed the terrestrial ionospheric cutoff (defined by the peak
ionospheric plasma frequency), which is typically 3--10~MHz.
\cite{fdz99} showed that $\nu_c$ exceeded the \objectname[]{Earth's}
ionospheric cutoff ($\sim 3$~MHz) for some planets known at the time
of the publication of their paper, with the planet around
\objectname[HD]{HD~114762} having $\nu_c \approx 1000$~MHz.

Just as the statistical variations in the solar-system planetary
magnetic moments contribute to variations in the radiometric Bode's
Law, there is also a statistical spread in the exponents and/or
coefficients for the emission frequency.  Compared to the radiated
power predicted by equation~(\ref{eqn:rbl}), the statistical
variations in the predicted emission frequency are relatively more
important.  The radiated power levels can be exponentially modified by
the solar-wind loading whereas the emission frequency is not.  In
order to account for the statistical variations in estimates of the
magnetic moment derived from Blackett's Law, we assume that the actual
emission frequency could be larger or smaller by a factor of~3 than
the nominal value of equation~(\ref{eqn:fc}).  (See also Appendix~A of
\citealt{fdz99}.)

The predicted flux density of an extrasolar planet is then 
\begin{equation}
S = \frac{P}{\Delta\nu\Omega D^2}.
\label{eqn:fluxdensity}
\end{equation}
\cite{fdz99} applied equations~(\ref{eqn:blackett})--(\ref{eqn:fluxdensity}) to
the nine then-known extrasolar planets.  The distances to the host
stars are known.  The planetary mass and distance to its parent star
are measured or constrained from the optical observations.  Assuming
an emission cone of~$4\pi$~sr, $\Delta\nu = \nu_c/2$, and a rotation
rate comparable to that of \objectname[]{Jupiter}, they derived the
resulting flux densities.  

  We shall do the same but for the currently
much larger sample.

\section{The Extrasolar Planet Census}\label{sec:census}

Since the initial work by \cite{fdz99}, the extrasolar planetary
census has grown by over an order of magnitude.  
We used the census of extrasolar planets contained in ``The Extrasolar
Planets Encyclopaedia'' \citep{s03}.
We used only the set of reasonably secure detections of planets around
main-sequence stars as of~2003 July~1.  The total number of planets
examined was 118 in~102 planetary systems.

The required quantities for applying equation~(\ref{eqn:rbl}) are the
planet's mass, distance from its primary, and rotation rate.  At the
time we extracted our list of planets from ``The Extrasolar Planets
Encyclopaedia,'' almost all of the known planets had been found via the
radial velocity method.  This method yields a lower limit to the
planet's mass (i.e., $M\sin i$).

Strictly, with the exception of the planets orbiting
\objectname[OGLE]{OGLE-TR-56} and \objectname[HD]{HD~209458}, for
which transits have been detected, the flux density levels that we
calculate are lower limits.  If the emitted radiation is nearly
isotropic and assuming that the inclinations of the orbits are random,
the typical error introduced is only a factor of~2.
There have been attempts to set upper limits on planetary masses,
e.g., using tidal constraints \citep{t00}.  We have not made use of
these upper limits because in many cases they introduce additional
uncertainties based on the characteristics of the host star (e.g., its
age).

The radial velocity method also determines the semi-major axis and
eccentricity of the planet's orbit.  We use the semi-major axis as the
distance~$d$ in equation~(\ref{eqn:rbl}).  For those planets in
eccentric orbits, we might expect the radio emission to be periodic
with the planet's orbital period, increasing near periastron and
decreasing near apastron.  The most extreme eccentricities in our
sample are approximately 0.7 (\objectname[HIP]{HIP~75458},
\objectname[HD]{HD~222582}, and \objectname[HD]{HD~2039}).  We
therefore expect a range of possible flux densities of possibly 20.
For a planet with a more typical eccentricity, $e \approx 0.1$, the
range is only a factor of~1.5 or so.

We have no information on the rotation rates of these planets.  We
have therefore assumed that those planets with semi-major axes larger
than 0.1~AU have rotation rates equal to that of \objectname[]{Jupiter} (10~hr),
while those with semi-major axes less than 0.1~AU are tidally-locked
with rotation rates equal to their orbital periods \citep{gs66,t00}.

For tidally-locked planets we have also assumed a radius of~1.25~$R_J$ 
in calculating the characteristic frequency.  This value is both
predicted based on the stellar insolation of such planets
\citep{gbhls96} and is close to the observed value for the planet
orbiting \objectname[HD]{HD~209458} \citep{cbng02}.

\begin{figure}
\epsscale{0.67}
\rotatebox{-90}{\plotone{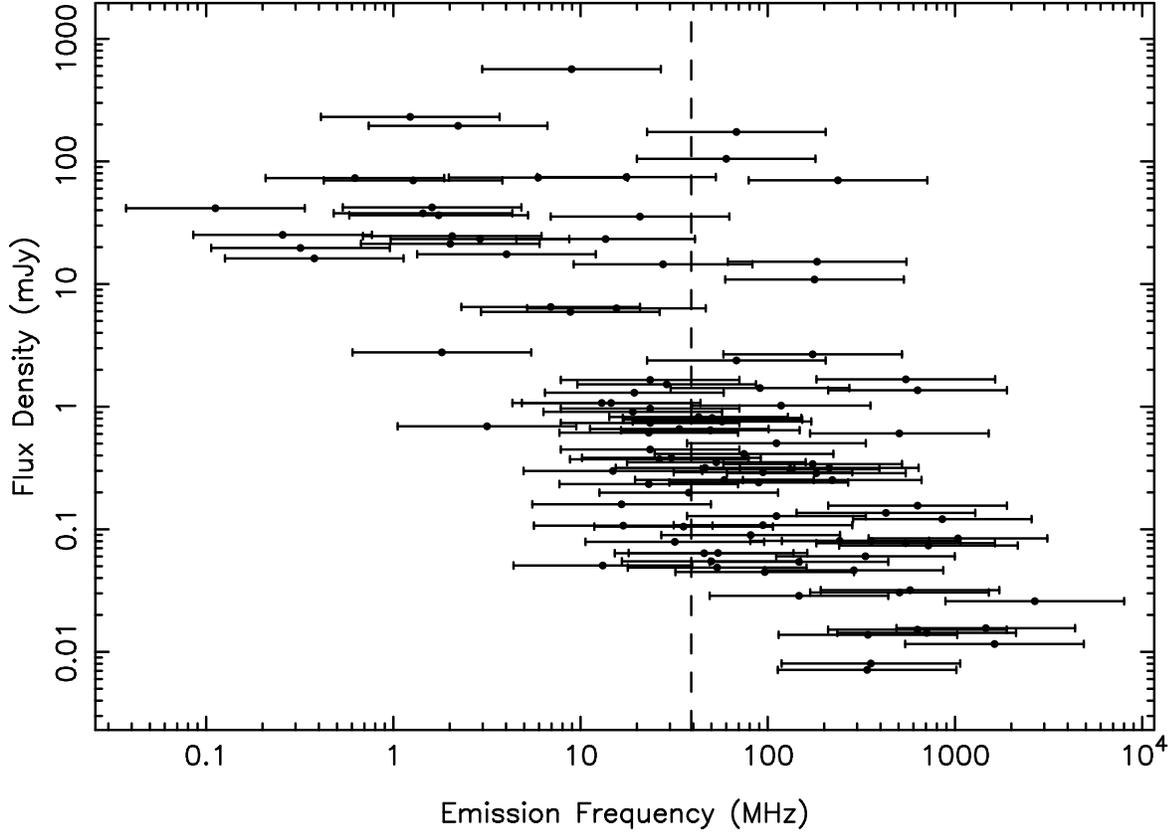}}
\caption[]{The predicted ``burst'' flux densities for 106 known extrasolar
planets vs.\ the characteristic emission frequency based on the
radiometric Bode's Law and Blackett's Law, equations~(\ref{eqn:rbl})--(\ref{eqn:fluxdensity}).  The horizontal bars indicate the assumed ranges
for the emission frequencies, allowing the statistical variations from
Blackett's Law in the estimated planetary magnetic moments.  The
expected \emph{burst} flux densities are obtained by assuming that
increases of roughly a factor of 100 can be obtained by larger stellar
wind loading of the planet's magnetosphere.  The vertical dashed line
indicates the approximate cutoff frequency for
\protect\objectname[]{Jupiter}.}
\label{fig:census}
\end{figure}

Table~\ref{tab:list} presents the full list of planets in our sample
along with the expected emission frequencies and flux density levels.
Figure~\ref{fig:census} presents the expected flux densities vs.\ the
emission frequency in a graphical form.  Anticipating our analysis of
\S\ref{sec:observe}, we follow the convention of \cite{fdz99} and
increase the predicted flux density levels by a factor of~100 in order
to account for variability driven by the incident stellar wind.

\begin{deluxetable}{lccccc}
\tablecaption{Exoplanet Radio Emission\label{tab:list}}
\tablewidth{0pc}
\tabletypesize{\small}
\tablehead{
 \colhead{Star} & \colhead{$\omega/\omega_J$} & 
 \colhead{$\log P_{\mathrm{rad}}$} & \colhead{$\mu$} &
 \colhead{$\nu_c$} & \colhead{$\log S$} \\
 & \colhead{ } & \colhead{(W)} & \colhead{(G\,R${}_J$)} &
 \colhead{(MHz)} & \colhead{(Jy)}}
\startdata

\objectname[]{OGLE-TR-56} & 0.3  &  16.95 &    2.6 &     14.8 & $-4.3$\\
\objectname[]{HD~73256}   & 0.2  &  16.77 &    3.7 &     20.8 & $-1.4$\\
\objectname[]{HD~83443}   & 0.1  &  15.79 &    0.3 &      1.4 & $-1.4$\\
\objectname[]{HD~46375}   & 0.1  &  15.48 &    0.1 &      0.6 & $-1.1$\\
\objectname[]{HD~179949}  & 0.1  &  16.28 &    1.1 &      5.9 & $-1.1$\\
\\
\objectname[]{HD~187123~b} & 0.1  &  15.88 &    0.4 &      2.1 & $-1.6$ \\
\objectname[]{HD~209458}  & 0.1  &  15.95 &    0.5 &      2.9 & $-1.6$ \\
\objectname[]{HD~75289}   & 0.1  &  15.65 &    0.2 &      1.3 & $-1.2$ \\
\objectname[]{BD~$-10\arcdeg$~3166} & 0.1  &  15.73 &    0.3 &      1.6 & \nodata \\
\objectname[]{HD~76700}   & 0.1  &  15.13 &    0.1 &      0.3 & $-1.7$\\
\objectname[]{$\tau$~Boo} & 0.1  &  16.93 &   10.7 &     59.9 & $-1.0$\\
\\
\objectname[]{51~Peg}     & 0.1  &  15.57 &    0.2 &      1.2 & $-0.6$\\
\objectname[]{HD~49674}   & 0.1  &  14.66 &    0.0 &      0.1 & $-1.4$\\
\objectname[]{$\upsilon$~And~b} & 0.09 &  15.67 &    0.4 &      2.2 & $-0.7$ \\
\objectname[]{$\upsilon$~And~c} & 1.0  &  15.25 &   12.1 &     68.0 & $-2.6$ \\
\objectname[]{$\upsilon$~And~d} & 1.0  &  14.87 &   38.0 &    213   & $-3.5$ \\
\\
\objectname[]{HD~168746}  & 0.07 &  14.86 &    0.0 &      0.3 & $-1.6$\ \\
\objectname[]{HD~217107~b} & 0.06 &  15.76 &    0.7 &      4.0 & $-1.8$ \\
\objectname[]{HD~68988}   & 0.07 &  16.02 &    1.6 &      8.8 & $-2.2$\\
\objectname[]{HD~162020}  & 0.05 &  17.06 &   31.7 &    178   & $-2.0$\\
\objectname[]{HD~130322}  & 0.04 &  15.36 &    0.4 &      2.0 & $-1.7$\\
\\
\objectname[]{HD~108147}  & 0.04 &  14.67 &    0.1 &      0.4 & $-1.8$\\
\objectname[]{Gl~86}      &    1 &  17.08 &   42.3 &    237   & $-1.2$\\
\objectname[]{55~Cnc~b}   &    1 &  16.15 &    3.1 &     17.6 & $-1.1$\\
\objectname[]{55~Cnc~c}   &    1 &  14.84 &    0.3 &      1.7 & $-1.4$\\
\objectname[]{55~Cnc~d}   &    1 &  14.32 &   43.2 &    242   & $-4.1$\\
\\					 			     
\objectname[]{HD~38529~b} &    1 &  16.03 &    2.8 &     15.5 & $-2.2$\\
\objectname[]{HD~38529~c} &    1 &  15.31 &  290.4 &   1626   & $-4.9$\\
\objectname[]{HD~195019}  &    1 &  16.83 &   32.8 &    184   & $-1.8$\\
\objectname[]{HD~6434}    &    1 &  15.64 &    1.2 &      6.9 & $-2.2$\\
\objectname[]{HD~192263}  &    1 &  15.88 &    2.4 &     13.6 & $-1.6$\\
\\					 			 
\objectname[]{Gl~876~b}   &    1 &  16.20 &   12.1 &     68.0 & $-0.8$\\
\objectname[]{Gl~876~c}   &    1 &  15.83 &    1.6 &      8.9 & $-0.2$\\
\objectname[]{$\rho$~CrB} &    1 &  15.82 &    4.9 &     27.6 & $-1.8$\\
\objectname[]{HD~74156~b} &    1 &  15.90 &    8.8 &     49.4 & $-3.2$\\
\objectname[]{HD~74156~c} &    1 &  15.05 &  120.7 &    676   & $-5.2$\\
\objectname[]{HD~3651}    &    1 &  14.69 &    0.3 &      1.6 & $-1.4$ \\
\\					 
\objectname[]{HD~168443~b} &    1 &  16.75 &  112.8 &    631   & $-2.9$ \\
\objectname[]{HD~168443~c} &    1 &  15.66 &  476.7 &   2670   & $-4.6$ \\
\objectname[]{HD~121504}  &    1 &  15.47 &    3.5 &     19.4 & $-2.9$\\
\objectname[]{HD~178911B} &    1 &  16.60 &   90.1 &    504.4 & $-3.2$\\
\objectname[]{HD~16141}   &    1 &  14.59 &    0.3 &      1.8 & $-2.6$\\
\\
\objectname[]{70~Vir}     &    1 &  16.42 &   97.5 &    546   & $-2.8$\\
\objectname[]{HD~80606}   &    1 &  16.03 &   32.4 &    182   & $-3.5$\\
\objectname[]{HD~219542B} &    1 &  14.59 &    0.6 &      3.2 & $-3.2$\\
\objectname[]{HD~216770}  &    1 &  15.08 &    2.3 &     13.0 & $-3.0$\\
\objectname[]{GJ~3021}    &    1 &  15.94 &   31.0 &    174   & $-2.6$\\
\\
\objectname[]{HD~52265}   &    1 &  15.31 &    5.1 &     28.8 & $-2.8$\\
\objectname[]{HD~37124~b} &    1 &  15.01 &    2.6 &     14.6 & $-3.0$\\
\objectname[]{HD~37124~c} &    1 &  14.22 &    5.7 &     31.9 & $-4.1$\\
\objectname[]{HD~73526}   &    1 &  15.67 &   26.2 &    147   & $-4.3$\\
\objectname[]{HD~8574}    &    1 &  15.40 &   16.0 &     89.5 & $-3.6$\\
\\					 			
\objectname[]{HD~104985}  &    1 &  15.98 &   90.3 &    505   & $-4.5$\\
\objectname[]{HD~134987}  &    1 &  15.18 &    9.0 &     50.4 & $-3.1$ \\
\objectname[]{HD~169830~b} &    1 &  15.51 &   23.5 &    132   & $-3.5$ \\
\objectname[]{HD~169830~c} &    1 &  14.53 &   17.2 &     96.3 & $-4.3$ \\
\objectname[]{HD~40979}   &    1 &  15.59 &   31.0 &    174   & $-3.5$ \\
\\					 
\objectname[]{HD~150706}  &    1 &  14.89 &    4.2 &     23.5 & $-3.1$\\
\objectname[]{HD~12661~b} &    1 &  15.36 &   16.8 &     94.3 & $-3.5$\\
\objectname[]{HD~12661~c} &    1 &  14.35 &    8.9 &     49.9 & $-4.3$\\
\objectname[]{HD~89744}   &    1 &  15.98 &  112.8 &    631   & $-3.8$\\
\objectname[]{HR~810}     &    1 &  15.27 &   16.2 &     90.9 & $-2.8$\\
\\					 			 
\objectname[]{HD~92788}   &    1 &  15.56 &   39.5 &    221   & $-3.6$\\
\objectname[]{HD~142}     &    1 &  14.76 &    4.2 &     23.5 & $-3.0$\\
\objectname[]{HD~177830}  &    1 &  14.89 &    6.3 &     35.5 & $-4.0$\\
\objectname[]{HD~128311}  &    1 &  15.30 &   21.0 &    118   & $-3.0$\\
\objectname[]{HD~28185}   &    1 &  15.73 &   76.4 &    428   & $-3.9$\\
\\					 			 
\objectname[]{HD~108874}  &    1 &  14.99 &    9.7 &     54.2 & $-4.2$\\
\objectname[]{HD~142415}  &    1 &  15.02 &   10.5 &     58.6 & $-3.6$\\
\objectname[]{HD~4203}    &    1 &  14.97 &    9.6 &     53.6 & $-4.3$\\
\objectname[]{HD~210277}  &    1 &  14.81 &    6.0 &     33.7 & $-3.2$\\
\objectname[]{HD~82943~b} &    1 &  14.93 &    9.5 &     53.1 & $-3.5$\\
\objectname[]{HD~82943~c} &    1 &  14.89 &    3.4 &     19.0 & $-3.0$\\
\\
\objectname[]{HD~27442}   &    1 &  14.84 &    7.6 &     42.7 & $-3.1$\\
\objectname[]{HD~114783}  &    1 &  14.61 &    4.1 &     23.1 & $-3.2$\\
\objectname[]{HD~20367}   &    1 &  14.63 &    4.7 &     26.3 & $-3.4$\\
\objectname[]{HD~147513}  &    1 &  14.59 &    4.2 &     23.5 & $-2.8$\\
\\
\objectname[]{HD~19994}   &    1 &  14.97 &   13.3 &     74.7 & $-3.4$\\
\objectname[]{HD~65216}   &    1 &  14.73 &    6.8 &     37.8 & $-3.7$\\
\objectname[]{HD~41004A}  &    1 &  15.04 &   16.8 &     94.3 & $-4.0$\\
\objectname[]{HIP~75458}  &    1 &  15.79 &  152.8 &    856   & $-3.9$\\
\objectname[]{HD~222582}  &    1 &  15.48 &   63.7 &    357   & $-4.1$\\
\\
\objectname[]{HD~160691}  &    1 &  14.78 &   10.2 &     57.0 & $-3.1$\\
\objectname[]{HD~23079}   &    1 &  15.01 &   19.9 &    111   & $-3.9$\\
\objectname[]{HD~141937}  &    1 &  15.77 &  185.3 &   1038   & $-4.1$\\
\objectname[]{HD~47536}   &    1 &  15.34 &   60.6 &    339   & $-5.1$\\
\\
\objectname[]{HD~114386}  &    1 &  14.41 &    4.1 &     23.1 & $-3.6$\\
\objectname[]{HD~4208}    &    1 &  14.26 &    3.0 &     16.6 & $-3.8$\\
\objectname[]{16~Cyg~B}   &    1 &  14.61 &    8.3 &     46.2 & $-3.5$\\
\objectname[]{HD~10697}   &    1 &  15.36 &   97.3 &    545   & $-4.1$ \\
\objectname[]{$\gamma$~Cep} &    1 &  14.52 &    9.1 &     50.9 & $-3.1$ \\
\\					 
\objectname[]{HD~213240}  &    1 &  15.12 &   51.5 &    289   & $-4.3$\\
\objectname[]{HD~111232}  &    1 &  15.43 &  128.8 &    722   & $-4.1$\\
\objectname[]{HD~114729}  &    1 &  14.12 &    3.0 &     16.9 & $-4.0$\\
\objectname[]{47~UMa~b}   &    1 &  14.77 &   19.9 &    111   & $-3.3$\\
\objectname[]{47~UMa~c}   &    1 &  13.67 &    2.7 &     14.9 & $-3.5$\\
\\			       	 				     
\objectname[]{HD~10647}   &    1 &  14.32 &    5.5 &     30.6 & $-3.4$\\
\objectname[]{HD~2039}    &    1 &  15.14 &   63.5 &    355   & $-5.1$\\
\objectname[]{HD~190228}  &    1 &  15.09 &   61.2 &    343   & $-4.9$\\
\objectname[]{HD~50554}   &    1 &  15.06 &   59.4 &    333   & $-4.2$\\
\objectname[]{HD~136118}  &    1 &  15.57 &  260.9 &   1461   & $-4.8$\\
\\			       	 				     
\objectname[]{HD~196050}  &    1 &  14.75 &   26.2 &    147   & $-4.5$\\
\objectname[]{HD~30177}   &    1 &  15.26 &  126.1 &    706   & $-4.8$\\
\objectname[]{HD~106252}  &    1 &  15.19 &  102.8 &    576   & $-4.5$\\
\objectname[]{HD~216435}  &    1 &  14.29 &    8.2 &     45.7 & $-4.2$\\
\objectname[]{HD~216437}  &    1 &  14.49 &   14.5 &     81.0 & $-4.0$\\
\\			       	 				       
\objectname[]{HD~23596}   &    1 &  15.19 &  112.5 &    630   & $-4.8$ \\
\objectname[]{14~Her}     &    1 &  14.94 &   59.4 &    333   & $-3.9$ \\
\objectname[]{HD~72659}   &    1 &  14.47 &   20.0 &    112   & $-4.8$ \\
\objectname[]{$\epsilon$~Eri} &    1 &  13.83 &    3.3 &     18.3 & $-2.2$ \\
\\					 
\objectname[]{HD~39091}   &    1 &  15.26 &  204.8 &   1147   & $-4.2$\\
\objectname[]{HD~33636}   &    1 &  15.39 &  183.4 &   1027   & $-4.3$\\
\objectname[]{Gl~777A}    &    1 &  13.82 &    6.8 &     37.8 & $-3.9$\\
\enddata
\tablecomments{Planets are arranged roughly in order of increasing
semi-major axis.  The predicted flux densities are the burst values.}
\end{deluxetable}

The trend in Figure~\ref{fig:census} of increasing emission frequency
and decreasing flux density is real and reflects two effects.  First,
the lower envelope reflects a selection effect.  A low flux density
and small emission frequency results from a low-mass planet in a large
orbit.  These planets cannot be detected with the current detection
methodology.  Second, the upper envelope reflects the well-known
deficit of planets with both large masses and small semi-major axes.
Even if high-mass planets with close orbits did exist, however, they
probably would be tidally locked (as we have assumed here).  The
rotation rate also determines the strength of the magnetic field
(equation~\ref{eqn:fc}), so tidally-locked planets probably do not
radiate at high emission frequencies.  However, \cite{z03} have
suggested that such ``hot Jupiters'' may radiate by the conversion of
stellar magnetic pressure (which is large in close to the star) into
electromotive forces, currents and radio emission.

We have assumed that the objects discussed here (Table~\ref{tab:list})
are in fact planets.  Using Hipparcos astrometric data, \cite{hbg01}
argue that the inclinations of the orbits of these objects imply that
$\sin i$ is small.  A small $\sin i$ would imply that the derived
masses of these objects would be larger, so that their magnetic
moments would be larger, but also that their distances from their
stars would be larger, so that the stellar wind loading of the
magnetosphere would be less.  The similar dependences for~$M$ and~$d$
in equation~(\ref{eqn:rbl}) mean that this impact of small $\sin i$
would nearly cancel.  However, if $\sin i$ is small enough, as
\cite{hbg01} suggest for a number of objects, they might no longer be
classified as planets but as brown dwarfs.  \cite{bergeretal01}
detected radio emission from one brown dwarf, indicating that it is
likely to be a magnetic body as well.  If so, the radiometric Bode's
Law and our analysis might still apply to them.

We now discuss selected objects in greater detail.
\begin{description}
\item[OGLE-TR-56] The planet orbiting this star was found from its
transits of its parent star \citep{ktjs03}.  In many respects it is
not much different than many of the other ``close-in'' planets.  Its
flux density is roughly 3 orders of magnitude lower, though, because
of its much greater distance (1500~pc vs.\ a more typical 10~pc).
This result is likely to be a general result as future planetary
searches move deeper into the Galaxy:  The detection of extrasolar
planetary radio emission is most likely to come from planets orbiting
nearby stars.

\item[HD~179949] A possible magnetic interaction between the planet
and the star has been detected for this star \citep{swb03}.

\item[$\tau$~Boo, HD~209458] These stars show \ion{Ca}{2}~H and~K
activity, similar to that of \objectname[HD]{HD~179949}.  However, a
clear link between the planetary orbital period and the modulation of
the \ion{Ca}{2}~H and~K emission has not been established yet \citep{swb03}.

\item[HD~187123] This star may have (at least) two planets orbiting
it \citep{vmba00}.  We provide a prediction for only the inner planet.
The outer planet has a period much longer than 3~yr.  Consistent with
other planets having relatively long periods ($d \sim 2~AU$), we
would expect the flux density from this second planet to be no more
than about $10^{-5}$~Jy.

\item[BD~$-10\arcdeg$~3166] This star has a planet orbiting it, but
the distance to the star is unknown, though \cite{bgmfha00} suggest a
distance of less than 200~pc.  If that is the case, its flux density
is probably larger than $10^{-3}$~Jy.

\item[$\upsilon$~And] There is uncertainty regarding the mass of the
outer planet, with \cite{bmfbcknn99} finding 4.6~M${}_J$ while
\cite{mztv99} find approximately 10~M${}_J$.  We have used the lower
value, realizing that this factor of two difference results in only a
modest change in both $\nu_c$ and $S$.

This star also shows \ion{Ca}{2}~H and~K
activity, similar to that of \objectname[HD]{HD~179949}.  However, a
clear link between the planetary orbital period and the modulation of
the \ion{Ca}{2}~H and~K emission has not been established yet \citep{swb03}.

\item[HD~217107] This star may have (at least) two planets orbiting
it \citep{fmbvfa01}.  We provide a prediction for only the inner planet.
The outer planet has a period much longer than 3~yr.  Consistent with
other planets having relatively long periods ($d \sim 2~AU$), we
would expect the flux density from this second planet to be no more
than about $10^{-5}$~Jy.

\item[HD~74156] The mass of the outer planet orbiting this star is
uncertain (in addition to the inclination factor).  The cited lower
limit is 7.5~M${}_J$.  A larger value would lead to both a higher
$\nu_c$ and~$S$.

\item[HD~92788] Estimates of the mass of the planet orbiting this star
differ by 5\% \citep{umq00,fmbvfa01}.  We have used the arithmetic
mean of the masses and semi-major axis in estimating the flux
density.

\item[HD~160691] The existing radial velocity data indicate, though do
not provide a compelling case, for the presence of an outer planet
\citep{jbmtpmc02}.  If this second planet does exist, its predicted
magnetic moment and radio emission would be similar to those of
\objectname[HD]{HD~216435}.

\item[$\epsilon$~Eri] \cite{qt02} claim the presence of an outer
planet based on the structure of the dust disk surrounding this star.
Even if such a planet exists, its large semi-major axis ($\approx
40$~AU) would imply that its radio emission would be similar to that
of \objectname[]{Neptune}, almost certainly at a frequency well below the
\objectname[]{Earth's} ionospheric cutoff.

\item[HD~192263] The existence of a planet around this star was
questioned on the basis of photometric variations with the same period
as the radial velocity variations \citep{hdb03}.  \cite{santosetal03}
present more recent radial velocity measurements, which appear to
confirm the presence of the planet.

\item[HD~47536] The mass of the star is uncertain and in the range
1--3~M${}_\odot$ \citep{setiawanetal01}.  We have used the lower limit
to obtain the mass of the planet.  Adopting a larger mass would imply
a larger predicted flux density, by a factor of as much as 4.

The star itself is a giant (spectral class K1III).  As the radiometric
Bode's Law was developed for planets orbiting a main-sequence star,
whether it can be applied to planets orbiting giant stars is not
clear.

\item[HD~114762] This object does not appear in Table~\ref{tab:list},
even though \cite{fdz99} made a prediction for its radio emission,
because it has since been re-classified as a brown dwarf \citep{s03}.
Radio emission may still be generated from such bodies, though
\citep{bergeretal01}.
\end{description}

\section{VLA 74~MHz Observations: An Initial
	Comparison}\label{sec:observe}

The model presented represents a zeroth-order estimate under the
assumption that a solar analog is valid. To confirm or refute the
model some initial confrontation with observations is warranted, even
if the observations cannot be obtained ``optimally'' in some sense.
For this purpose we utilize the VLA Low-frequency Sky Survey (VLSS),
an effort to survey the northern sky using the VLA's 74~MHz system
\citep{kassimetal03}.  We make use of this survey for two reasons.
First, at a frequency of~74~MHz, the survey frequency is within a
factor of~2 of the cutoff frequency of the radio emission of \objectname[]{Jupiter}
($\simeq 40$~MHz) and within a factor of~2 of the predicted emission
frequencies for nearly one-third of the current census.

Second, in producing Table~\ref{tab:list} we have reported ``burst''
flux densities, which are a factor of~100 larger than the median
values.  Our rationale for doing so is that we occasionally may detect
the bursts from extrasolar planets.  The VLSS observations are a
series of overlapping pointing centers on the sky with each pointing
center observed in five, 15-min.\ ``snapshots.''  While not as
sensitive as the targeted observations of \cite{bdl00}, the VLSS
offers the potential for obtaining multiple observations on a
substantial fraction of the extrasolar planet census.  (The limitation
on the number of stars that can be observed by the VLSS is its
southern declination limit of~$-30\arcdeg$.  Thus, approximately
one-third of the stars in the current census cannot be observed.)  By
observing a large number of stars, the VLSS has the potential for
observing a rare, large burst.

\begin{figure}
\epsscale{1}
\plotone{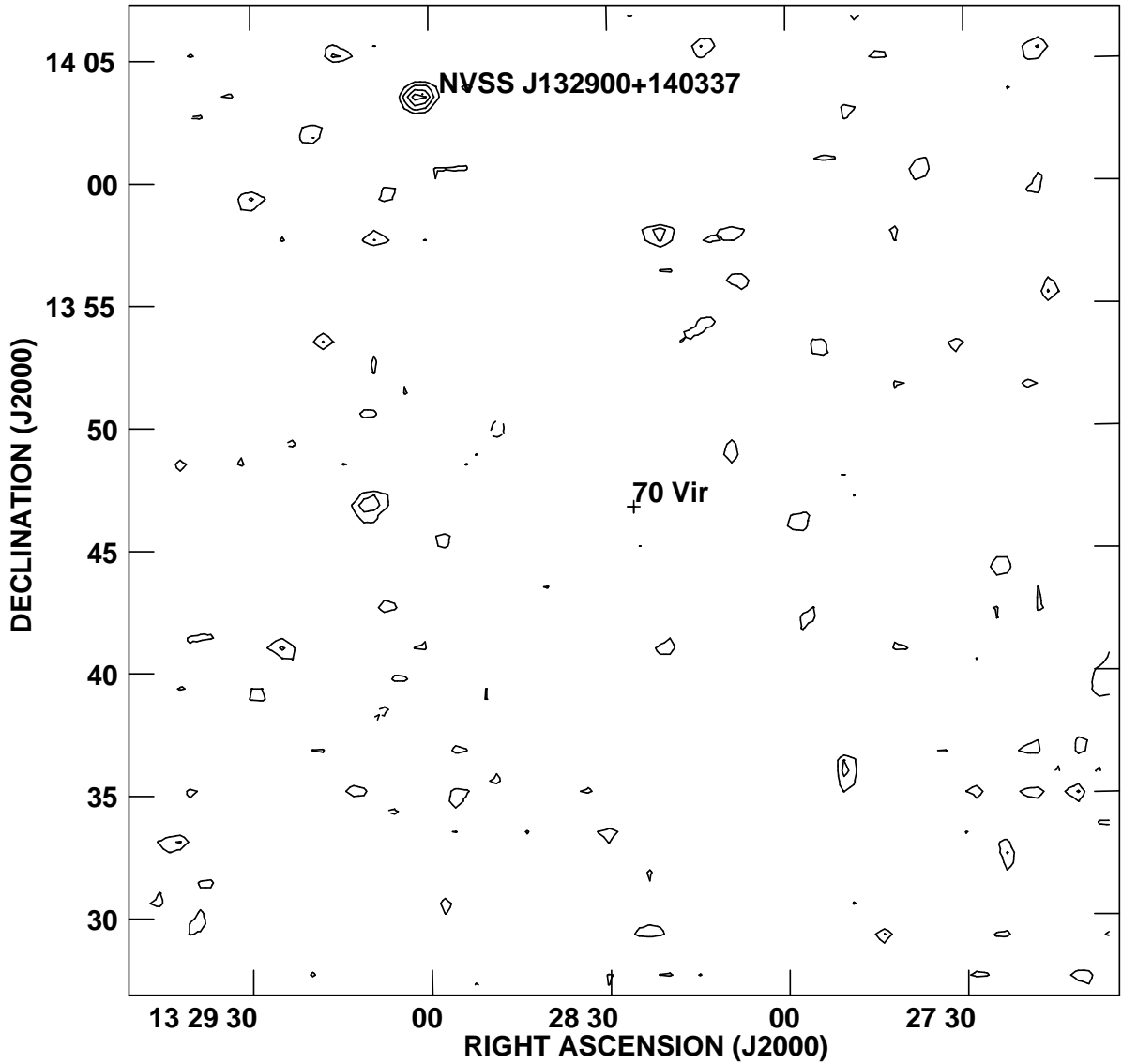}
\vspace{-1.5cm}
\caption[]{The field around 70~Vir at~74~MHz, illustrating the typical
field observed in the \hbox{VLSS}.  The noise level is 0.13~Jy~beam${}^{-1}$, with a
25\arcsec\ $\times$ 25\arcsec\ beam.  The contours are
0.13~Jy~beam${}^{-1}$ $\times$ $-3$, 2, 3, 4, 5, 7.07, and~10.  The
position of 70~Vir is marked by a cross.  Also indicated is the radio
source \objectname[NVSS]{NVSS~J132900$+$140337}, likely to be an
extragalactic radio source.}
\label{fig:observe}
\end{figure}

Presently, the VLSS is 10\% complete, having surveyed a a 0.9~sr
region.  As an illustration of its future utility as well as to
confront the models with what limited observations may be available,
we have examined the locations of the known extrasolar planets within
the current survey region.  Figure~\ref{fig:observe} shows the field
around \objectname[]{70~Vir} as a typical example.

In the current census, 5 stars fall within the region surveyed thus
far.  In no case have we detected emission at the location of an
extrasolar planet.  Table~\ref{tab:observe} summarizes the upper
limits we derive; we reproduce the predicted emission frequencies and
``burst'' flux densities for convenience as well.  We obtain the upper
limits by determining the local noise level ($\sigma$), which is
typically near 100~mJy.  Radio sources in common to the 74~MHz VLSS
and the 1400~MHz NVSS are used to register the positions of sources
detected in VLSS, and experience shows that the reference frames at
the two frequencies agree to within approximately 5\arcsec.  The radio
reference frame provided by the NVSS itself should be aligned to
within~40~milliarcseconds of the optical reference frame.  Thus, the
locations of the stars hosting extrasolar planets are known to a much
better precision than the beam (point-spread function) of the VLA
at~74~MHz (80\arcsec).  As a result, we have adopted a signal-to-noise
threshold of~2.5$\sigma$ in setting upper limits.

\begin{deluxetable}{lcccc}
\tablecaption{74~MHz VLSS Constraints on Extrasolar Planetary Radio Emission\label{tab:observe}}
\tablewidth{0pc}
\tablehead{
\colhead{Star} & \colhead{$\nu_c$} & \colhead{$S_{\mathrm{model}}$} & \colhead{$S_{\mathrm{observe}}$} \\
               & \colhead{(MHz)} & \colhead{(mJy)}                  & \colhead{(mJy)}}
\startdata
\objectname[HD]{HD~104985} & 505 & 0.03 & 220 \\
		     		 
\\		     		 
		     		 
\objectname[HD]{HD~108874} &   54 & 0.06 & 325 \\
\objectname[]{70~Vir}     &  546 &  1.6 & 325 \\
\objectname[]{$\tau$~Boo} &   60 &  100 & 275 \\
\objectname[]{$\rho$~CrB} &   28 &   16 & 218 \\

\enddata
\tablecomments{The model flux densities are the ``burst'' flux
densities, obtained by multiplying the median levels predicted by
equation~(\ref{eqn:rbl}) by a factor of~100.  The upper limits given
by $S_{\mathrm{observe}}$ are 2.5$\sigma$ in the VLSS 74~MHz images.}
\end{deluxetable}

Of the 5 extrasolar planets within the currently surveyed region of
the VLSS, two have predicted emission frequencies well above the
survey frequency of~74~MHz.  On the basis of the model, we would not
expect to see emission at the location of these stars.  We emphasize,
though, that one of our purposes in comparing the VLSS observations to
the model predictions is an effort to confront the model with whatever
observations are available.  

For the remaining 3 extrasolar planets, their emission frequencies are
all within nearly a factor of~2 of~74~MHz.  In these cases we can
explain the absence of emission as due to the low flux densities
expected, though \objectname[]{$\tau$~Boo} is a possible exception.
For \objectname[]{$\tau$~Boo} the observed upper limit is within a
factor of only 3 of the expected ``burst'' emission levels.  As
\cite{fdz99} discussed, \objectname[]{$\tau$~Boo} remains a
potentially useful target for further tests.  (See also
\citealt*{fldbz03} for further observations of \objectname[]{$\tau$~Boo}.)

\section{Discussions and Conclusions}\label{sec:conclude}

We have used the extrasolar planet census, as of~2003 July~1, and
scaling laws developed in our solar system to predict the emission
frequencies and flux densities for radio emission from extrasolar
planets.  Our work presents an update and extension of the initial
work of \cite{fdz99}.  Figure~\ref{fig:census} and
Table~\ref{tab:list} present our primary results.  Selection effects
due to the current planet detection methodology produce an apparent
deficit of planets with low flux densities and small emission
frequencies.  There is also a deficit of planets with high flux
densities and large emission frequencies due to the well-known deficit
of planets with both large masses and small semi-major axes.

The most likely planets to be detected are those orbiting nearby
stars.  Thus, the planet orbiting \objectname[OGLE]{OGLE-TR-56} is
similar to many of the other ``close-in'' planets, but its flux
density is roughly 3 orders of magnitude lower than most of the other
planets in the census because of its much greater distance (1500~pc
vs.\ a more typical 10~pc).  This result is likely to be a general
result as future planetary searches move deeper into the Galaxy.

We have also used the initial observations of the 74~MHz VLA
Low-frequency Sky Survey (VLSS) to place constraints on the emission
from a small number of planets.  In no case do we detect any emission,
and the typical upper limit is approximately 0.3~Jy.  For most of the
planets, the predicted flux density is orders of magnitude lower than
the upper limit, but the upper limit for the planet orbiting
\objectname[]{$\tau$~Boo} is within a factor of~2 of the predicted
``burst'' flux density.

Most of the known extrasolar planets have expected emission
frequencies below~1000~MHz.  Indeed, if equation~(\ref{eqn:fc}) is
correct, we do not expect extrasolar planets to emit strongly
above~1000~MHz because their magnetic moments are not large enough.
Thus, efforts to detect the radio emission from extrasolar planets
must focus on the low-frequency capabilities of current or future
radio telescopes.  

In designing an experiment to detect extrasolar planets, the potential
for bursts must be taken into account.  The sensitivity of a radio
telescope improves with increasing integration time as $t^{-1/2}$.
Longer integration times improve the sensitivity but at the cost of
``diluting'' any bursts.  For a burst of flux density~$S_b$ and
duration~$\Delta t_b$, its average flux density in an integration
time~$t$ is $S_b(\Delta t_b/t)$, i.e., the burst is diluted with time
as $t^{-1}$.  To the extent that extrasolar planet radio emission is
``bursty'' (as is observed for the solar system planets), multiple
short observations are a better strategy than long integrations, as
are employed usually in radio astronomy.

We shall now estimate the sensitivity of various radio telescopes for
detecting extrasolar planets.  In estimating radio telescope
sensitivities, particularly for aperture synthesis instruments, it is
common to use long integration times ($\sim 8$~hr).  In contrast, the
magnetospheres of the solar system planets respond to solar wind
variations on time scales of order 1~hr.  We shall use 15~min.\ as an
integration time both to emphasize that long integration times may not
be appropriate for estimating sensitivities and to allow for the
possibility that shorter magnetosphere response times may occur.

The Very Large Array has been used already in attempts to detect radio
emission from extrasolar planets \citep{bdl00,fldbz03}, and as we
illustrate here, its on-going 74~MHz northern sky survey provides an
excellent opportunity to continue testing these models.
Below~1000~MHz, the VLA operates at~74 and~330~MHz.  Its nominal
sensitivity at these two frequencies is roughly
$200\,\mathrm{mJy}/\!\sqrt{t/15\,\mathrm{min.}}$ and
$1.5\,\mathrm{mJy}/\!\sqrt{t/15\,\mathrm{min.}}$ over bandwidths of~1.5
and~3~MHz, respectively.  As Figure~\ref{fig:census} demonstrates,
only if the radiometric Bode's Law vastly underestimates the radiated
power levels will the VLA be able to detect radio emissions from
extrasolar planets, even taking into account the possibility of strong
bursts.

The Giant Metrewave Radio Telescope (GMRT) operates at frequencies
of~150, 235, 330, and~610~MHz.  Over this frequency range, its nominal
sensitivity varies somewhat but is roughly
$0.2\,\mathrm{mJy}/\!\sqrt{t/15\,\mathrm{min.}}$ in a 4-MHz bandwidth.
Although larger bandwidths (and therefore improved sensitivities) are
technically possible with the GMRT, we use a 4-MHz bandwidth in an
attempt to strike a balance between improved sensitivity and avoiding
terrestrially-generated radio interference.  As
Figure~\ref{fig:census} shows, most extrasolar planets are below the
detection limit of the GMRT, unless they produce strong bursts (factor
of roughly 100 increase in flux density).

The Low Frequency Array (LOFAR) is a radio telescope in the design and
development phase that is expected to operate in the frequency range
10--240~MHz.  This frequency range encompasses the predicted emission
frequencies of roughly two-thirds of the current census.  The
sensitivity of LOFAR will be highly frequency- and
direction-dependent, due to the effect on the system temperature from
the Galactic synchrotron emission.  For directions away from the
Galactic plane, the design goals specify a sensitivity, in a 15-min.\
integration with a 4-MHz bandwidth, around~2~mJy at the lower
frequencies and increasing to around~1~mJy at the higher frequencies.
There are a number of potential candidates for observations with
\hbox{LOFAR}.  Perhaps most promising are \objectname[]{$\tau$~Boo},
\objectname[Gl]{Gl~876}, and \objectname[Gl]{Gl~86} as these three
objects may be detectable even in the absence of any flux density
increase due to stellar wind variability.  Also of interest are
\objectname[HD]{HD~162020} and \objectname[HD]{HD~195019}, which may
require only moderate increases (factor of~10) due to stellar wind
variability to become detectable.

The Square Kilometer Array (SKA) is a next-generation radio telescope
that is expected to operate above~150~MHz.  Its current design goals
suggest that its sensitivity would be approximately 1~$\mu$Jy in a
15-min. integration.  This sensitivity is sufficient that it should be
capable of detecting the radio emissions from the most massive
extrasolar planets, without relying upon bursts to enhance the
emission levels, and a substantial fraction of the current census if
their burst levels are comparable to what we find here.  The efficacy
of the SKA in detecting the radio emission from extrasolar planets
depends upon two aspects of its design, the lowest frequency at which
it operates and its collecting area.  First, if the potential SKA
frequency range were to be increased to extend below~150~MHz, the
number of planets it potentially could detect would also increase.
Second, as stated currently, the SKA sensitivity is specified in terms
of a ratio between its effective collecting area and system
temperature, $A_{\mathrm{eff}}/T_{\mathrm{sys}}$.  This quantity is
taken to be constant.  As $T_{\mathrm{sys}}$ increases below~1000~MHz
due to Galactic synchrotron emission, this design goal requires that
$A_{\mathrm{eff}}$ also increase below~1000~MHz so as to keep the
ratio constant.  If this design goal is relaxed so that
$A_{\mathrm{eff}}/T_{\mathrm{sys}}$ decreases below~1000~MHz, the
likelihood of the SKA detecting the radio emission from extrasolar
planets also decreases.

All of our discussions assume that the radio emissions from these
planets would be driven solely by the stellar wind loading of the
planets' magnetospheres.  If there are large, close-in satellites,
like \objectname[]{Io} at \objectname[]{Jupiter}, higher emission
frequencies and flux densities could result.  Also, for planets in
close to their parent stars, magnetic pressures that are large in the
interior heliosphere may drive radio emission \citep{z03}. Future
modeling efforts include integrating the current census with these hot
Jupiter emission models.

\acknowledgements
We thank J.~Schneider for his efforts in producing and maintaining
``The Extrasolar Planet Encyclopedia,'' without which this project
would have been considerably more arduous.  Basic research in radio
astronomy at the NRL is supported by the Office of Naval Research.

\end{document}